\documentclass[final,5p,times,twocolumn,numbers]{elsarticle}
\usepackage{amsmath}
\usepackage{amssymb}
\usepackage{lipsum}
\usepackage{booktabs}
\usepackage[utf8]{inputenc}
\usepackage[T1]{fontenc}
\usepackage{float}
\usepackage{caption}
\usepackage{subcaption}

\usepackage{booktabs, makecell, tabularx}
\usepackage{natbib}
\usepackage{booktabs, makecell, tabularx}

\usepackage{siunitx}%
\usepackage{adjustbox}
\usepackage{physics} %Vasile

\def\tsc#1{\csdef{#1}{\textsc{\lowercase{#1}}\xspace}}
\tsc{WGM}
\tsc{QE}
\journal{Physics Letters B}

\begin{document}

\begin{frontmatter}



\title{Novel way of evaluating $g_A$ quenching in $\beta^+$/EC decays : Introducing the Branching-Ratio Method (BRM)}


\author[first]{Aagrah Agnihotri}
\ead{aagrah.a.agnihotri@jyu.fi}
\author[first,second]{Jouni Suhonen}
\ead{jouni.t.suhonen@jyu.fi}
\affiliation[first]{organization={University of Jyv\"askyl\"a},Department of Physics
            addressline={P.O. Box 35}, 
            postcode={FI-40014},
            city={Jyv\"askyl\"a},
            state={},
            country={Finland}}  
\affiliation[second]{organization={International Centre for Advanced Training and Research in Physics (CIFRA)},%Department and Organization
            addressline={P.O. Box MG12}, 
             postcode={077125},
            city={Bucharest-M{\u a}gurele}, 
            state={},
            country={Romania}}

\begin{abstract}
The novel Branching-Ratio Method (BRM) for the determination of effective value $g_{\rm A}^{\rm eff}$ of the weak axial coupling $g_{\rm A}$ for forbidden non-unique (FNU) $\beta^+$/electron-capture(EC) decays is introduced. The method offers new possibilities for testing the fitness of nuclear Hamiltonians in modeling the physics of complex $\beta^+$/EC decays. The constraint of simultaneously reproducing the branching to $\beta^+$ and EC transitions offers an additional constraint in tackling the problem of $g_{\rm A}^{\rm eff}$ determination. 
%The true power of this method is in determining the g$_{\rm A}^{\rm eff}$ of forbidden non-unique decays. 
In the BRM the ambiguity in the choice of the values of $g_{\rm A}^{\rm eff}$ and s-NME (small relativistic vector nuclear matrix element) is lifted when constraints of the branching ratios of $\beta^+$ and EC decays are applied. As an example, in the present work we apply BRM to the case of second FNU $\beta^+$/EC decay of $^{59}$Ni. 
%whose spectral shape is insensitive $g_{\rm A}^{\rm eff}$ and s-NME values. g$_{\rm A}^{\rm eff}$ determination for this decay would remain intractable without the intervention of the branching ratio method. 
This decay is treated with three different nuclear shell-model (NSM) Hamiltonians demonstrating the effects of different nuclear-structure aspects in application of the BRM.  
%Spectral Shape Studies are performed for the first time for $\beta^+/EC$ forbidden non-unique decays. $g_A^{eff}$ dependence of $\beta^+/EC$ branching ratio offers a novel tool for the determination of $g_A^{eff}$ for forbidden non-unique $\beta^+/EC$. In addition to the prospects offered by the enhanced Spectral Shape Method, the additional constraint on $g_A^{eff}$ by the $\beta^+/EC$ branching ratio opens a new avenue of research concerning the physics of forbidden non-unique $\beta$ decay and determination of $g_A^{eff}$.
\end{abstract}




%\tableofcontents

\end{frontmatter}

\section{Introduction}
\label{introduction}
Determination of the effective value of the weak axial-vector coupling $g_{\rm A}$, i.e. $g_{\rm A}^{\rm eff}$, is an important problem that must be resolved in the study of weak-interaction nuclear processes \cite{avcreview,Suhonen2019,Neutrinonuclearres,PhysRevC.96.055501}. In particular, the study of neutrinoless double beta ($0\nu\beta\beta$) decay has implications concerning the foundational questions in physics \cite{PhysRevC.96.055501,RevModPhys.92.045007,RevModPhys.95.025002,Engel_2017} and its understanding in terms of $g_{\rm A}^{\rm eff}$ in phenomenological nuclear models is of high priority. Importance of $g_{\rm A}^{\rm eff}$ determination for $0\nu\beta\beta$ in one of the key incentives for studying the phenomenology of $g_{\rm A}^{\rm eff}$ for all weak-interaction nuclear processes. Further incentives for the investigation of the $g_{\rm A}^{\rm eff}$ phenomenology include the accurate modeling of astrophysical processes  \cite{RevModPhys.75.819,Langanke_2021}, resolving the reactor anti-neutrino flux and "bump" anomalies \cite{Hayen2019a,Hayen2019b,Ramalho2022,universe7100360,ZHANG2024104106} and in determination of neutrino mass using low $Q$-value EC decays \cite{NatPhys18,PhysRevLett.124.222503,Gastaldo2017,GE2022137226,PhysRevC.106.015501,PhysRevLett.127.272301,PhysRevC.107.015504,Redshaw2023,Ruotsalainen:2024gxu}. Practically, solving the aforementioned broad questions in physics involves understanding the two classes of weak-interaction nuclear processes, namely $\beta^-$ and $\beta^+$/electron-capture (EC) decays, as this understanding is a necessary stepping stone in resolving such questions. 
%An example would be understanding the g$_{\rm A}^{\rm eff}$ phenomenology for $\beta^+$/EC decays would be key to predict the sensitivity experiments aiming to measure $\beta^+$/EC type $0\nu\beta\beta$ decays \cite{PhysRevC.96.055501,RevModPhys.92.045007,RevModPhys.95.025002} Therefore, the importance of  g$_{\rm A}^{\rm eff}$ phenomenology manifests itself on all levels of analysis directly or indirectly, in multiple areas of nuclear, astro-, and particle physics. 

The characterization of the phenomenology of $g_{\rm A}^{\rm eff}$ has been a long-standing mystery, and only in recent years this mystery has been resolved for a restricted set of light nuclei by P. Gysbers \textit{et al.} \cite{quenchingresolved}. This work makes it clear that the need to use a renormalized value of $g_{\rm A}$ (or actually a renormalized spin-isospin operator) is the manifestation of, e.g., missing nuclear many-body correlations in the computed $\beta$-decay nuclear matrix elements (NMEs); a consequence part and parcel of using approximations, i.e. adopting nuclear models with phenomenological many-body Hamiltonians, instead of the complete \textit{ab initio} one. The virtual impossibility of solving the nuclear many-body problem exactly in all nuclear mass regions makes the use of limited phenomenological nuclear models inevitable. Determination of $g_{\rm A}^{\rm eff}$ is an effective and practical way to renormalize the involved spin-isospin operator 
%quantify the effects of lacking nuclear structure contents in the NMEs and study such trends 
across the nuclide chart for the phenomenological nuclear models \cite{avcreview}. 
%Such quantification helps predict insufficiencies in computations, and one can systematically overcome the absence of perfect nuclear Hamiltonians, and work towards resolving important open questions in physics. 

Studies have been done for understanding $g_{\rm A}^{\rm eff}$ phenomenology in allowed and first-forbidden unique $\beta$ decays \cite{avcreview,Neutrinonuclearres,Suhonen2019,SURENDER2024169772,PhysRevC.108.014327}, but considerable attention has lately been paid to evaluations of  $g_{\rm A}^{\rm eff}$ in forbidden non-unique (FNU) $\beta^-$ decays, in particular in terms of $\beta$-electron spectral shapes  \cite{introssm,ssm,PhysRevC.100.065805,24na36clssm,59fe60fessm,PhysRevC.109.014326,PhysRevC.109.034321,2020135092,Kostensalo2021,PhysRevLett.129.232502,PhysRevLett.133.122501,Kostensalo2023,Kostensalo2024,Bandac2024,Ramalho2024,Paulsen2024}.
%with major efforts made in the recent years (by nuclear theory group led by Jouni Suhonen at JYU) \cite{avcreview, Neutrinonuclearres,Suhonen2019}(add forbidden non-unique references). 
The degree to which different flavors of $\beta^-$ decays are treated in the literature, and understanding of $g_{\rm A}^{\rm eff}$ therein, depend on the complexity and difficulty of modeling a given decay type. Some work has been lately done to understand the FNU EC decays in terms of $g_{\rm A}^{\rm eff}$ in \cite{PhysRevLett.133.232501}, but
%Allowed and forbidden $\beta^-$ , EC and  $\beta^+$/EC variants have been studied to varying degrees \cite{avcreview,Neutrinonuclearres,Suhonen2019,SURENDER2024169772}. 
no general and consistent treatment for combined FNU $\beta^+$ and EC decays exists in the present literature since they are the most complex of all $\beta$ decays, and therefore aspects of $g_{\rm A}^{\rm eff}$ on this front are a terra incognita. 

This letter is an attempt to pave the way for the exploration of the less discussed physics of FNU $\beta^+$/EC decays. For this, we introduce here the novel Branching-Ratio Method (BRM) for $g_{\rm A}^{\rm eff}$ determination that is applicable to all $\beta^+$/EC decays. For demonstration of the power of BRM, we apply it to the 2$^{\rm nd}$ FNU $\beta^+$/EC ground-state-to-ground-state transition $^{59}\textrm{Ni}(3/2^-)\to\,^{59}\textrm{Co}(7/2^-)$. The BRM for $g_{\rm A}^{\rm eff}$ determination is different from mainstream $\beta^-$ methods, like the spectrum-shape method (SSM) \cite{introssm,ssm}, its enhanced version \cite{24na36clssm,59fe60fessm} and the spectral moment method (SMM) \cite{Kostensalo2023,Kostensalo2024}, in that it can be consistently applied to all $\beta^+$/EC decays. BRM uses the relative branching ratios between the  $\beta^+$ and EC sub-branches of a $\beta^+$/EC transition to determine $g_{\rm A}^{\rm eff}$. This method is particularly helpful for FNU decays 
%two exp. constraints are necessary to determine g$_{\rm A}^{\rm eff}$
, but in principle equally applicable to allowed and forbidden unique (FU) decays. The essence of $g_{\rm A}^{\rm eff}$ determination for the FNU $\beta$ decays involves treating the value of $g_{\rm A}^{\rm eff}$ as a free parameter and fitting its value to reproduce the branching ratios/(partial) half-lives and spectral shapes simultaneously. Such fitting suffices for modeling all features of allowed and FU $\beta^-$ decays as well, but without the spectral-shape aspect since these decays have universal spectral shapes, independent of $g_{\rm A}^{\rm eff}$. In BRM the difficulty of reproducing the $\beta^+$ and EC branching ratios simultaneously is overcome 
%upon the infrequent occurrence of theoretical $g_{\rm A}^{\rm eff}$-dependent spectral shapes, 
with the aid of an additional free parameter, namely the small relativistic vector NME (sNME), used already in the enhanced SSM \cite{24na36clssm,59fe60fessm} and SMM \cite{Kostensalo2023,Kostensalo2024}. An interesting new aspect of the sNME is that a $\beta^-$ spectral shape may also depend strongly on the sNME itself \cite{24na36clssm,59fe60fessm,PhysRevC.109.014326,PhysRevC.109.034321}. Hence, two experimental constraints are always needed for FNU decays to determine $g_{\rm A}^{\rm eff}$ and sNME simultaneously. 

BRM is a powerful tool to test the suitability of nuclear Hamiltonias for description of weak-interaction processes, in particular the $\beta$ decays. Here $g_{\rm A}^{\rm eff}$ and sNME can be determined by the total branching of a $\beta^+$/EC transition and the relative branching of $\beta^+$ and EC in this transition. A severe check of the consistency of this fitting is the comparison of the resulting positron spectral shape with the measured one. This is the more rewarding if the spectral shape would depend on $g_{\rm A}^{\rm eff}$ and/or sNME. Such a double check would qualify a phenomenological or an \textit{ab initio} Hamiltonian with considerable confidence. Such exposure of the imperfections of the adopted Hamiltonians will help improve the quality of nuclear-structure models.

\section{Theory}
%%\label{}
\subsection{Theory of $\beta$ decay} 
The non-relativistic limit of the theory of $\beta$ decays is well established and a detailed account is presented in \cite{Behrens1982ElectronRW}. A more concise and practical exposition of the treatment of forbidden $\beta^+$/EC decays is given in 
\cite{ECformalism}. The half-life value for $\beta^+$/EC decays, $t^{\beta^{+}/\rm EC}_{1/2}$, is given as
\begin{equation}
\label{betaechf}
    t^{\beta^{+}/\rm EC}_{1/2}=\frac{\kappa}{\tilde{C}^{\beta^+}+\tilde{C}^{\mathrm{EC}}},
\end{equation}
where $\kappa$ takes the value of 6289 s. $\tilde{C}^{\beta^+}$ and $\tilde{C}^{\mathrm{EC}}$ are the integrated shape functions of $\beta^+$ and EC decays, respectively.
The integrated shape function of forbidden non-unique $\beta^-$/$\beta^+$  decay is given as 
\begin{equation}
\label{shape} 
\tilde{C}^{\mp}=\int_{1}^{\ w_{0}}\!C^{\mp}(w_{\mathrm{e}})p w_{\mathrm{e}}(w_{0}-w_{\mathrm{e}})^{2}F_{0}(\pm Z_f,w_{\mathrm{e}})\,d w_{\mathrm{e}},
\end{equation}
where $w_{e}$ is the total energy of the emitted electron/positron and p is the electron/positron momentum, scaled by the electron rest mass. Here $Z_{f}$ is the charge number of the daughter nucleus, and $F_{0}(Z_{f},w_{e})$ is the Fermi function.
The integrated shape function for EC decay can be written as \cite{ECformalism}
\begin{equation}
\label{intSF}
    {\Tilde{C}}^{\rm EC}=\frac{\pi}{2}\sum_{x=1s,2s}n_x\beta_x^2(p_{\nu_x}/m_ec)^2C(p_{\nu_x})\,,
\end{equation}
when considering the leading channels, namely K capture (1s) and L$_{\mathrm{1}}$ capture (2s). Here $n_x$ are relative occupancies for the respective orbitals, $\beta_x$ are the electron Coulomb amplitudes, and $p_{\nu_x}$ is the momentum of the neutrino when the electron is captured from the atomic orbital $x$ \cite{ECformalism}. The shape factor $C(p_{\nu_x})$ used in Eq.~(\ref{intSF}), containing complicated combinations of phase-space factors and nuclear matrix elements (NMEs), is given for forbidden EC transitions in detail in \cite{ECformalism}. 

According to Eq.\ref{betaechf}, the (partial) half-life $t_{1/2}^{\beta^+/\rm EC}$ of the $\beta^+$/EC decay transition is related to the partial half-lives $t_{1/2}^{\beta^+}$ and $t_{1/2}^{\rm EC}$ of its $\beta^+$ and EC sub-branches, respectively, as:

\begin{equation}
\label{halflife}
    \frac{1}{t_{1/2}^{\beta^+/\rm EC}}= \frac{1}{t_{1/2}^{\beta^+}}+\frac{1}{t_{1/2}^{\rm EC}} .
\end{equation}
In case of a $\beta^+$/EC transition between two nuclear states, the total branching ratio is given as:
\begin{equation}
\label{br}
    \rm BR_{\beta^+/ \rm EC} \%=\frac{T^{tot}_{1/2}}{t_{1/2}^{\beta^+/\rm EC}}\times100,
\end{equation}
with $T^{\rm tot}_{1/2}$ being the total $\beta^+$/EC half-life of the parent nucleus.
The branching ratios of $\beta^+$ and EC sub-branches of a given $\beta^+$/EC transition, with total half-life $t_{1/2}^{\beta^+/\rm EC}$, are given in a similar way as outlined in Eq.~(\ref{br}), thus being
\begin{equation}
\label{brbetaec}
    \rm BR_{\beta^+} \%=\frac{t_{1/2}^{\beta^+/\rm EC}}{t_{1/2}^{\beta^+}}\times100,\quad  \rm BR_{EC} \%=\frac{t_{1/2}^{\beta^+/\rm EC}}{t_{1/2}^{\rm EC}}\times100 .
\end{equation}
Finally, following Eq.~(\ref{brbetaec}), we define the relative branching ratio of the $\beta^+$ sub-branch with respect to the EC sub-branch for a given $\beta^+$/EC transition as 
\begin{equation}
\label{relbr}  
   \rm BR_{\beta^+: \rm EC} = \frac{\rm BR_{\beta^+}}{\rm BR_{EC}} = \frac{\rm t_{1/2}^{\rm EC}}{t_{1/2}^{\beta^+}} .
\end{equation}
Computed average $\beta$ energy, $\langle E_{\beta^{\mp}}\rangle$, for $\beta^{\mp}$ decays, can also be relevant when speculating the physical relevance of a $\beta$ spectral-shape prediction. One can compute $\langle E_{\beta^{\mp}}\rangle$ using the integrated shape function of Eq. (\ref{shape}) using the following relation:
\begin{equation}
\label{betaavg}
\langle E_{\beta^{\mp}}\rangle=\frac{\int_{0}^{E_0}E\times\tilde{C}^{\mp}\,dE}{\int_{0}^{E_0}\tilde{C}^{\mp}\,dE}
\end{equation}
where $E$ is the kinetic energy of the $\beta$ particle and $E_0$ is the $\beta$ end-point energy.

\subsection{Branching-Ratio Method (BRM) for $\beta^+$/EC decays}
\label{BRM}

As alluded to in the introduction, the Branching-Ratio Method (BRM) for $g_{\rm A}^{\rm eff}$ determination in a $\beta^+$/EC decay transition involves necessarily applying a measured relative branching ratio, defined by Eq.~(\ref{relbr}), as a constraint to theoretical computations and the fit of $g_{\rm A}^{\rm eff}$ as a free parameter. The basis of the BRM is the physical assumption that theoretical NMEs must reproduce simultaneously the partial half-life $t_{1/2}^{\beta^+/\rm EC}$ of the $\beta^+$/EC decay transition of interest (in case of just one possible transition this is the total half-life $T_{1/2}^{\rm tot}$ of the mother nucleus) and the relative branching of Eq.~(\ref{relbr}).
%$t_{1/2}^{\beta^+/\rm EC}$ of the $\beta^+$/EC branch and  $t_{1/2}^{\beta^+}$ and $t_{1/2}^{\rm EC}$ of its $\beta^+$ and EC sub-branches simultaneously. 
The $\beta^+$/EC decays are sensitive to imperfections/incompleteness of NME of phenomenological nuclear models and depend non-trivially on $g_{\rm A}^{\rm eff}$ (and sNME in case of FNU decays). Therefore, ${\rm BR}_{\beta^+: \rm EC}$ of Eq.~(\ref{relbr}) is a demanding experimental observable to reproduce theoretically; yet for exactly the same reason, it makes for an excellent test and standard for improving the quality of phenomenological nuclear models and Hamiltonians. 

It should be noted here that for a FNU $\beta^+$/EC transition the simultaneous BRM description of both the half-life of the $\beta^+/\rm EC$ transition of interest and the relative branching of Eq.~(\ref{relbr}) is highly non-trivial and can succeed or fail for a phenomenological Hamiltonian. For the FNU decays the sNME can be brought in as an additional parameter, as done in the enhanced SSM \cite{24na36clssm,59fe60fessm} and SMM \cite{Kostensalo2023,Kostensalo2024}. However, even this does not guarantee a successful application of the BRM, as shown below by our example Hamiltonians in the case of the FNU $\beta^+$/EC decay of $^{59}$Ni. A rough guess for the value of sNME is its CVC-based value, sNME$_{\rm CVC}$, obtained by relating the value of sNME to the value of a large vector NME, lNME, as described, e.g., in Refs.~\cite{Kostensalo2023,Kostensalo2024}. This value of sNME pertains, however, to an ideal many-body nuclear theory and thus serves only as a rough reference for the sNME values obtained in the BRM for the imperfect phenomenological nuclear Hamiltonians.  

Potential cross check of the BRM is the measurement of the $\beta^+$-decay positron spectrum. The related positron spectral shape gives an extra constraint for the BRM-extracted values of $g_{\rm A}^{\rm eff}$ and sNME, be it dependent or independent of $g_{\rm A}^{\rm eff}$. For a perfect model Hamiltonian all three constraints, namely the $\beta^+/\rm EC$ half-life and relative branching as also the positron spectral shape should be simultaneously reproduced by a unique pair of values $(g_{\rm A}^{\rm eff},\textrm{sNME})$. For an imperfect Hamiltonian the three experimental observables open up several ways to explore the physical content of the Hamiltonian and thus gain understanding for future improvements of it.
Hence, measurements of the positron spectral shapes are welcome for validation and improvement of Hamiltonias of both phenomenological and \textit{ab initio} nuclear models.

\section{Calculations and results}
%%\label{}

\subsection{Nuclear-Structure Calculations \label{sec:structure}}
\begin{figure}
    \centering
    \caption{Comparison of the experimental and GX1APN-, KB3GPN- and FPD6PN-computed level energies (in MeV) of $^{59}$Ni.}
    \includegraphics[width=\linewidth]{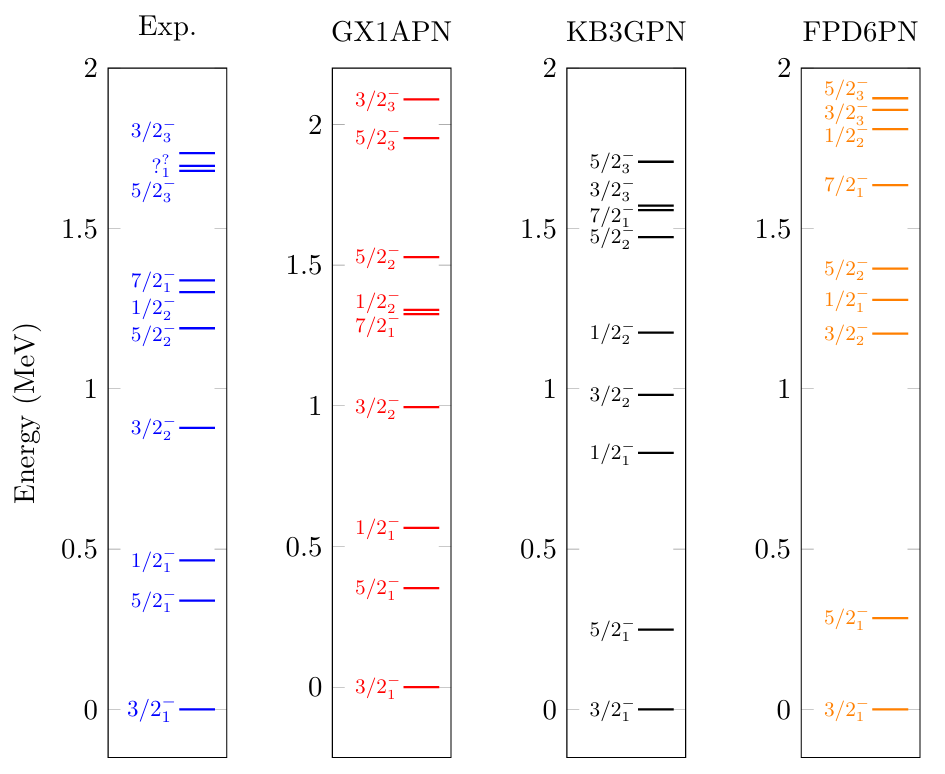}
    \label{fig:levels_59ni}
\end{figure}

\begin{figure}
    \centering
    \caption{Comparison of the experimental and GX1APN-, KB3GPN- and FPD6PN-computed level energies (in MeV) of $^{59}$Co.}
    \includegraphics[width=\linewidth]{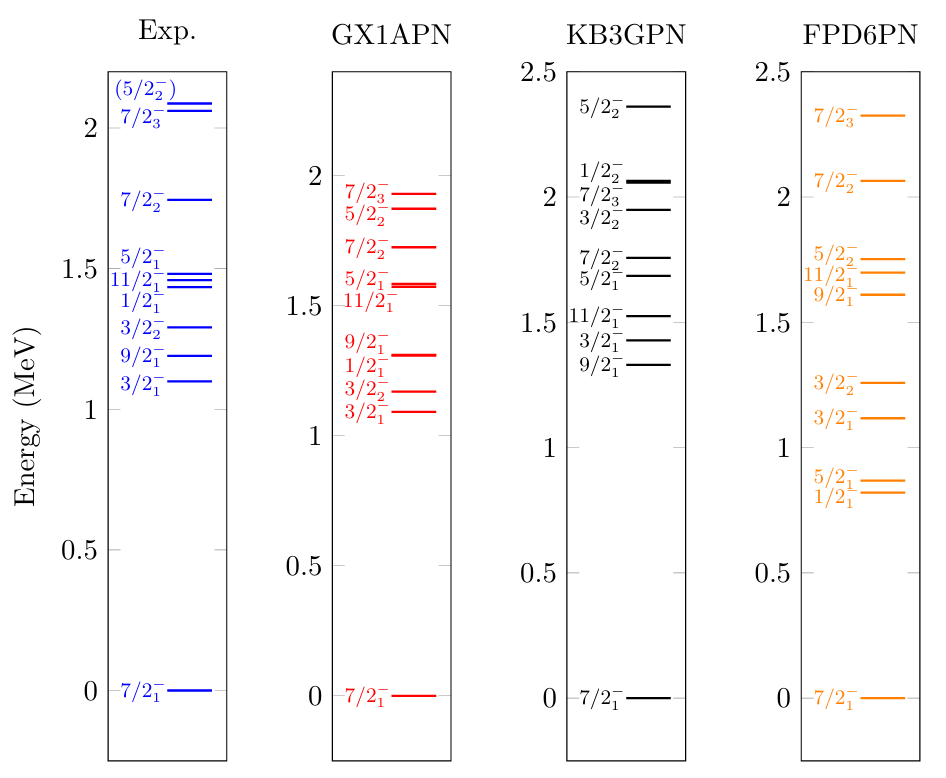}
    \label{fig:levels_59co}
\end{figure}

Nuclear wave functions for the NMEs and electromagnetic (EM) observables are computed using the NSM computer program NuShellX$@$MSU \cite{BROWN2014115}. For the parent and daughter nuclei three well-established nuclear Hamiltonians in the PFPN model space are used to draw out different 
nuclear-structure manifestations of the same $\beta^+$/EC decay transition through the lens of the BRM. The three Hamiltonians correspond to three interactions namely GX1APN, KB3GPN, and FPD6PN. The computed energy-level schemes for $^{\rm 59}$Ni (parent nucleus) and $^{\rm 59}$Co (daughter nucleus) are given in Fig.\ref{fig:levels_59ni} and Fig.\ref{fig:levels_59co}, respectively. In general, the experimental level energies are reproduced rather well for all interactions, with the GX1APN interaction giving the best and quite nice agreement between theory and experiment. In table \ref{tab:Electromagnetic_moments}, magnetic dipole and electric quadrupole moments of the ground states of $^{\rm 59}$Ni and $^{\rm 59}$Co are given for the three interactions. Experimental values of EM moments of the ground states are known only for $^{\rm 59}$Co, and again the GX1APN interaction gives the best estimates for the observables.

\begin{table}
\caption{Experimental and theoretical values of EM moments of the states involved in the $\beta^+$/EC decay of $^{\rm 59}$Ni. The magnetic dipole ($\mu$) and electric quadrupole ($Q$) moments are given in units of nuclear magnetons ($\mu_N/c$) and $e$barns, respectively. In the calculations, an effective charge 1.5$e$ (0.5$e$) for proton (neutron) and bare $g$ factors were used \cite{Suhonen2007}. Experimental values are taken from \cite{NNDC}.} 
\label{tab:Electromagnetic_moments}
\begin{tabular}{ccccc}
\toprule
 Interaction & $\mu_{\rm exp}$ & $\mu_{\rm theo}$ & $Q_{\rm exp}$ & $Q_{\rm theo}$ \\ \midrule
$^{59}_{28}$Ni$^{}_{31}$(3/2$^-_1$) & & & & \\ \midrule
GX1APN & - & -0.632 & - & +0.0662 \\ 
KB3GPN & - & -1.083 & - & -0.0163 \\
FPD6PN & - & -0.657 & - & -0.1486 \\ \midrule
$^{59}_{27}$Co$^{}_{32}$(7/2$^-_1$) & & & & \\ \midrule                               
GX1APN & +4.615(25) & +4.543 & +0.42(3) & +0.4407 \\ 
KB3GPN &  & +4.398 &  & +0.3946 \\
FPD6PN &  & +4.544 & & +0.4678 \\      
\bottomrule
\end{tabular}
\end{table}

\subsection{BRM for the Hamiltonians GX1APN, KB3GPN and FPD6PN}
\label{BRMfrHamiltonians}
As discussed in section \ref{BRM}, one can fit the free parameters to the experimental partial half-lives $t_{1/2}^{\beta^+/\rm EC}$, $t_{1/2}^{\beta^+}$ and $t_{1/2}^{\rm EC}$ through the  $\beta^+$/EC-decay half-life $t_{1/2}^{\beta^+/\rm EC}$ and the relative branching of Eq.~(\ref{relbr}). For the FNU $\beta^+$/EC decay ground-state-to-ground-state transition $^{59}\textrm{Ni}(3/2^-)\to\,^{59}\textrm{Co}(7/2^-)$ the fitting of $t_{1/2}^{\beta^+/\rm EC}$ will result in a pair of sNME values for each $g_{\rm A}^{\rm eff}$,
%and we label them as sNME$_{\beta^+/EC}$, sNME$_{\beta^+}$, and sNME$_{EC}$ resp. Also, each set of such sNME values will have two subsets 
labeled sNME$_{\beta^+/\rm EC}$(1) and sNME$_{\beta^+/\rm EC}$(2) due to the existence of two sNME solutions for a given $g_{\rm A}^{\rm eff}$ \cite{PhysRevLett.133.122501,Kostensalo2021,Kostensalo2023,Kostensalo2024,2020135092,PhysRevC.109.014326,PhysRevC.109.034321,Bandac2024,Ramalho2024}. The
 Q-value of the decay under study is 1073.00(19) keV \cite{NNDC}, with strong EC feeding and very weak $\beta^+$ feeding. The $\rm BR_{\beta^+}$ and $\rm BR_{\rm EC}$ are 3.7(12)$\times$10$^{-7}$ and 0.9999696(1) \cite{NNDC}, respectively, corresponding to $\rm t^{\beta^+}_{1/2}$=6.48$^{+2.15}_{-2.15}$$\times$10$^{18}$ s and $\rm t^{\rm EC}_{1/2}$=2.40$^{+0.16}_{-0.16}$$\times$10$^{12}$ s. Clearly, the relative branching ratio (\ref{relbr}) will practically be the same as $\rm BR_{\beta^+}$, hence we take its value to be 3.7(12)$\times$10$^{-7}$.
 Also, $\rm t^{\rm EC}_{1/2}$ and $\rm t^{\beta^+/\rm EC}_{1/2}$ are practically the same for the purposes of model calculations, therefore we take both values to be the same. 
 
 Given the above premises, we proceed to discuss the strategy to address the half-lives $t_{1/2}^{\beta^+/\rm EC}$ and $t_{1/2}^{\beta^+}$ and the relative branching BR$_{\beta^+: \rm EC} $ of Eq.~(\ref{relbr}), including their experimental error bands by using the GX1APN interaction. For this we define two independent sNME: sNME$_{\beta^+/\rm EC}$ and sNME$_{\beta^+}$. We fit g$_{\rm A}^{\rm eff}$ and sNME$_{\beta^+/\rm EC}$ to reproduce the experimental half-life $t_{1/2}^{\beta^+/\rm EC}$ with its error band. The resulting fitted values, sNME$_{\beta^+/\rm EC}$(1) and sNME$_{\beta^+/\rm EC}$(2), together with the resulting relative branchings, are displayed in table \ref{sNME-GX1APN} in rows 2 - 6. We can also fit g$_{\rm A}^{\rm eff}$ and sNME$_{\beta^+}$ to reproduce the experimental half-life $t_{1/2}^{\beta^+}$ with its error band. The corresponding fitted values, sNME$_{\beta^+}$(1) and sNME$_{\beta^+}$(2), together with the resulting relative branchings, are displayed in table \ref{sNME-GX1APN} in rows 8 - 12. 
 
\begin{figure}
    \centering
    \caption{Experimental mean values and error bands for the half-lives $\rm t^{\beta^+/\rm EC}_{1/2}$ (red) and $t_{1/2}^{\beta^+}$ (blue), and the resulting lower limits (ll), mean values (m) and upper limits (ul) for sNME$_{\beta^+/\rm EC}$ and sNME$_{\beta^+}$. The overlap region defines the physically relevant values of sNME and $g_{\rm A}^{\rm eff}$, as indicated by the horizontal and vertical bars centered at the mean values of the involved sNME(1) and sNME(2).}
    \includegraphics[width=\linewidth]{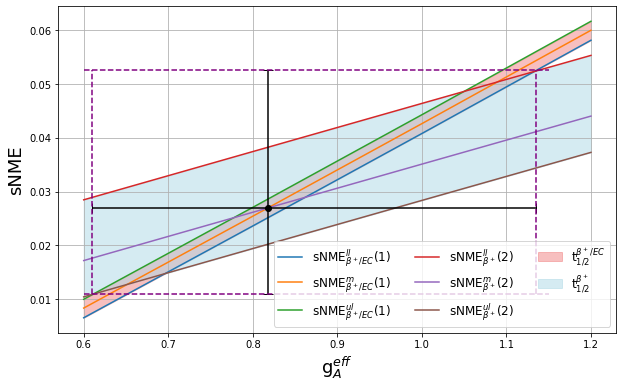}
    \label{fig:bounds}
\end{figure}

 First important thing that is observed from these results, is that the two sNME solutions, sNME$_X$(1) and sNME$_X$(2), $X=\beta^+/\textrm{EC},\beta^+$, always produce different orders of magnitude for the relative branching. This fact speaks for the utility of the relative BR in determining the physically relevant sNME value when compared with the experimental value. It is seen from these results that branching ratios of $\beta^+$ and EC sub-branches need not be simultaneously reproduced using the computed sNME$_{\beta^+/\rm EC}$. Existence or non-existence of sNME$_{\beta^+/\rm EC}$ for a given  $g_{\rm A}^{\rm eff}$ that simultaneously reproduces the relative branching is dependent on the adopted nuclear model. We see from the results in rows 2 - 6 that for the GX1APN interaction, the experimental relative BR is reached only for the sNME$_{\beta^+/\rm EC}$(1) subset for $g_{\rm A}^{\rm eff}=0.818$ and  sNME$_{\beta^+/\rm EC}(1)=0.02697^{+0.00163}_{-0.00182}$. Similarly, from the results in rows 8 - 12 we see that the experimental relative BR is reached only for the sNME$_{\beta^+}$(2) subset for $g_{\rm A}^{\rm eff}=0.818$ and  sNME$_{\beta^+}(2)=0.02697^{-0.00647}_{+0.01123}$. 
 The significant experimental uncertainty in the half-lives $\rm t^{\beta^+/\rm EC}_{1/2}$ and $t_{1/2}^{\beta^+}$, leads to a significant uncertainty in the values of $g_{\rm A}^{\rm eff}$ and sNME determined using the BRM. 
 The upper/lower bounds of the possible $g_{\rm A}^{\rm eff}$ and sNME values can be determined upon applying the constraints of upper/lower bounds of $t_{1/2}^{\beta^+}$ and $\rm t^{\beta^+/\rm EC}_{1/2}$ as shown in figure \ref{fig:bounds}. This yields $g_{\rm A}^{\rm eff}$$\in$[0.61,1.135] 
and sNME$\in$[0.011,0.0525] for the physically acceptable ranges of these parameters. Within these ranges the measured values of $\rm t^{\beta^+/\rm EC}_{1/2}$ and $t_{1/2}^{\beta^+}$ are simultaneously reproduced, the central values being $g_{\rm A}^{\rm eff}$=0.818 and $\textrm{sNME}=0.0270$, and the associated ranges being $g_{\rm A}^{\rm eff}=0.818^{+0.317}_{-0.208}$ and $\textrm{sNME}=0.0270^{+0.0255}_{-0.016}$. The latter values deviate notably from the CVC-determined value $\textrm{sNME}_{\rm CVC}=0.175$.
%In the cases when the relative branching ratios are not simultaneously reproduced, one has the option to fit s-NME to beta+ and EC branching ratios. But in this case, it is not necessary that total branching will be the same as the experimental branching ratios. The physical relevance of these branching ratios remains the subject of further investigation. In the scenario when the theoretically computed beta spectra are gA and s-NME sensitive, the physical significance of beta+ s-NME can be tested using the enhanced spectral shape method and spectral moments method. But again the effective gA determined this was and its relation to the gA deduced from BRM remains an open question. 

\begin{table}[H]
\caption{GX1APN-computed s-NME values, sNME$_{\beta^+/EC}$(1) and sNME$_{\beta^+/EC}$(2), columns 2 and 4, which reproduce the experimental total half-life (rows 2 - 6) or the $\beta^+$ half-life (rows 8 - 12) for a given $g^{\rm eff}_{\rm A}$ of the  2$^{\rm nd}$ FNU $\beta^+/ \rm EC$ decay transition of $^{59}$Ni. The respective relative BR, $\rm BR_{\beta^+: \rm EC}$, are also given in columns 3 and 5. The last row gives the CVC value of sNME.}
\centering
    \label{sNME-GX1APN}
    \resizebox{0.50\textwidth}{!}{%
    \begin{tabular}{ccccc}
    \toprule
        
        $g^{\rm eff}_{\rm A}$ & sNME$_{\beta^+/\rm EC}$(1) & $\rm BR_{\beta^+: \rm EC}$ & sNME$_{\beta^+/\rm EC}$(2) & $\rm BR_{\beta^+: \rm EC}$ \\ \midrule
        0.6  & 0.00839$^{+0.00163}_{-0.0018}$ & 2.52$^{+0.38}_{-0.37}$e-7 & 0.10703$^{-0.00163}_{+0.00179}$ & 2.87$^{+0.12}_{-0.12}$e-6 \\ 
        %0.7 & 0.01689 & 3.03e-7 & 0.11525 & 3.03e-6 \\
        0.8 & 0.02543$^{+0.00164}_{-0.00181}$ & 3.60$^{+0.50}_{-0.48}$e-7 & 0.12344$^{-0.00164}_{+0.00181}$ & 3.19$^{+0.14}_{-0.14}$e-6 \\
        0.818 & 0.02697$^{+0.00163}_{-0.00182}$ & 3.71$^{+0.50}_{-0.50}$e-7 & 0.12491$^{-0.00164}_{+0.00181}$ & 3.21$^{+0.14}_{-0.14}$e-6 \\
        %0.9 & 0.03400 & 4.22e-7 & 0.13160 & 3.35e-6 \\
        1.0 & 0.04260$^{+0.00165}_{-0.00183}$ & 4.89$^{+0.63}_{-0.61}$e-7 & 0.13972$^{-0.00165}_{+0.00183}$ & 3.51$^{+0.15}_{-0.16}$e-6 \\
        %1.1 & 0.05124 & 5.62e-7 & 0.14780 & 3.68e-6 \\
        1.2 & 0.05992$^{+0.00167}_{-0.00185}$ & 6.41$^{+0.78}_{-0.76}$e-7 & 0.15585$^{-0.00167}_{+0.00185}$ & 3.85$^{+0.17}_{-0.17}$e-6 \\
        %1.27 & 0.06602 & 7.01e-7 & 0.16146 & 3.97e-6 \\
         & sNME$_{\beta^+}$(1) & & sNME$_{\beta^+}$(2) & \\
        0.6  & -0.08301$^{+0.00673}_{-0.01127}$ & 4.70$^{-0.81}_{+1.33}$e-8 & 0.01720$^{-0.00673}_{+0.01127}$ & 5.39$^{-2.37}_{+9.37}$e-7 \\
        %0.7   & -0.07857 & 4.45e-8  & 0.02168 & 4.50e-7 \\
        0.8   & -0.07414$^{+0.00673}_{-0.01126}$ & 4.21$^{-0.74}_{+1.24}$e-8 & 0.02616$^{-0.00673}_{+0.01126}$ & 3.81$^{-1.58}_{+5.56}$e-7 \\
        0.818 & -0.07334$^{+0.00673}_{-0.01125}$ & 4.17$^{-0.73}_{+1.23}$e-8 & 0.02697$^{-0.00647}_{+0.01123}$ & 3.71$^{-1.54}_{+5.31}$e-7 \\
        %0.9   & -0.06969 & 4.00e-8 &  0.03063 & 3.26e-7 \\
        1.0   & -0.06524$^{+0.00673}_{-0.01126}$ & 3.80$^{-0.69}_{+1.15}$e-8 & 0.03509$^{-0.00673}_{+0.01126}$ & 2.82$^{-1.12}_{+3.61}$e-7 \\
        %1.1   & -0.06078 & 3.61e-8 & 0.03955 & 2.47e-7\\
        1.2   & -0.05632$^{+0.00673}_{-0.01126}$ & 3.44$^{-0.63}_{+1.07}$e-8 & 0.04400$^{-0.00673}_{+0.01126}$ & 2.17$^{-0.82}_{+2.50}$e-7\\
        %1.27  & -0.05320  & 3.33e-8 & 0.04711 & 2.00e-7\\
        & \multicolumn{2}{c}{sNME$_{\rm CVC}$=0.175}  & & \\
        \bottomrule
    \end{tabular}}
\end{table}

At this point it should be noted that the other two test Hamiltonians, KB3GPN and FPD6PN, could not perform successfully when applying the BRM: Neither one could reproduce simultaneously the experimental partial half-lives $t_{1/2}^{\beta^+/\rm EC}$ and $t_{1/2}^{\beta^+}$, i.e. no unique pair ($g_{\rm A}^{\rm eff}$,sNME) could be found to explain the central values of both half-lives. Hence, it can be said that GX1APN is the best interaction of the available three for the description of $\beta^+$/EC decay properties in this case. This is in line with the finding that also the electromagnetic moments were found to be best reproduced by this interaction in section \ref{sec:structure}.

\subsection{$\beta$ spectra}

\begin{figure}
\centering
\caption{Computed spectral-shape predictions for the three interactions under consideration.The labels lower limits (ll), mean values (m) and upper limits (ul) are for sNME values for a given $g_{\rm A}^{\rm eff}$ obtained upon fitting to uncertainty limits of the total half-life $t_{1/2}^{\beta^+/\rm EC}$. The spectral shape depicted in panel (b) is "universal" corresponding to all values of $g_{\rm A}^{\rm eff}$, sNME$_{\beta^+/\rm EC}(1)$ and sNME$_{\beta^+/\rm EC}(2)$ for the interactions GX1APN and FDP6PN, with $g_{\rm A}^{\rm eff}=1$ and sNME$_{\beta^+/\rm EC}(2)$ taken as a showcase.}
\label{shapes}
     \begin{subfigure}[b]{0.5\textwidth}
         \centering
         \caption{KB3GPN:spectra for (g$_{\rm A}^{\rm eff}$,sNME$_{\beta^+/EC}$(2))}
         \includegraphics[width=\textwidth]{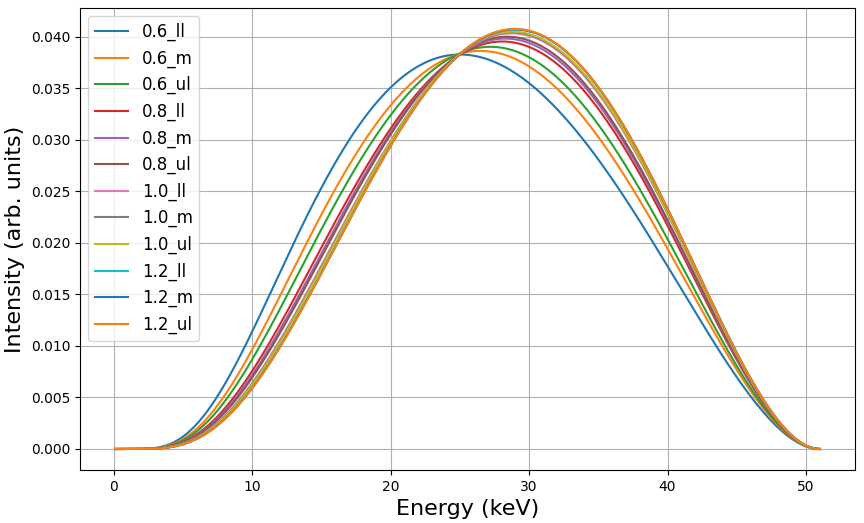}
         \label{KB3GPN}
     \end{subfigure}
     \label{fig:three_graphs}
     \begin{subfigure}[b]{0.5\textwidth}
         \centering
         \caption{GX1APN and FPD6PN:spectra for (g$_{\rm A}^{\rm eff}$=1.0,sNME$_{\beta^+/EC}$(2))}
         \includegraphics[width=\textwidth]{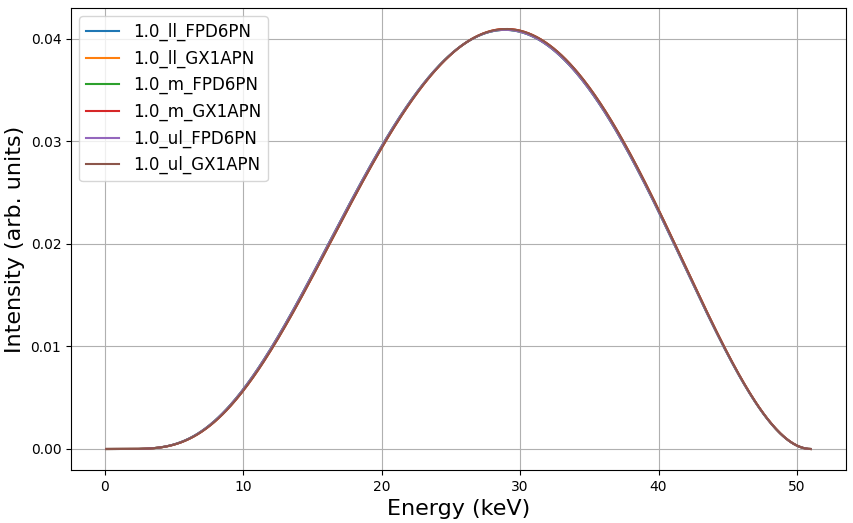}
         \label{GX1APN/FPD6PN}
     \end{subfigure}
\end{figure}

Out of twelve subsets of sNME values (four for each adopted Hamiltonian), only one subset, corresponding to sNME$_{\beta^+/\rm EC}$(2) for the KB3GPN interaction, shows $g_{\rm A}^{\rm eff}$/sNME dependence of the positron spectral shape, presented in Fig.~\ref{KB3GPN}. The spectra for the remaining eleven subsets are $g_{\rm A}^{\rm eff}$/sNME independent and practically have the same spectral shape, and we refer to this shape as the universal "default" shape. Therefore, a showcase spectral shape for the GX1APN and FPD6PN interactions is given for reference in Fig.~ \ref{GX1APN/FPD6PN}, corresponding to $g_{\rm A}^{\rm eff}$=1.0. In Table~\ref{avgbeta}, the ranges of average positron energies $\langle E_{\beta^+}\rangle$ of Eq.~(\ref{betaavg}), corresponding to the spectral shapes of Fig.~\ref{shapes}, are presented. Here each spectral shape corresponds to a unique $\langle E_{\beta^+}\rangle$.
%i.e. same spectral shape will have the same $\langle E_{\beta^+}\rangle$. The $g_{\rm A}^{\rm eff}$/sNME sensitive spectra in the case of KB3GPN interaction, can carry physically relevant geometry of spectra, given that nuclear structure calculations are not perfect, with exp. $t_{1/2}^{\beta^+}$ not being reproduced in this case (since exp. $t_{1/2}^{\beta^+/\rm EC}$ is reproduced), a potential consequence of renormalization of NMEs. 
The visible trend, as seen in Fig.~\ref{KB3GPN} and Table~\ref{avgbeta}, is that as $g_{\rm A}$ becomes less quenched, the spectral shape approaches the "default" shape presented in Fig.~ \ref{GX1APN/FPD6PN}. Therefore the primary prediction for the spectral shape would be the "default" shaped spectrum, valid for $g_{\rm A}^{\rm eff}=0.818^{+0.317}_{-0.208}$ and $\textrm{sNME}=0.0270^{+0.0255}_{-0.016}$ as determined in Sec. \ref{BRMfrHamiltonians} using the GX1APN interaction.

\begin{table}
\caption{ The average positron energies $\langle E_{\beta^+}\rangle$ corresponding to computed spectral shapes presented in Fig.~\ref{shapes}}
\centering
    \label{avgbeta}
    \resizebox{0.50\textwidth}{!}{%
    \begin{tabular}{ccc}
    \toprule
        KB3GPN & &  \\
        $g^{\rm eff}_{\rm A}$ & sNME$_{\beta^+/\rm EC}$(2) & $\langle E_{\beta^+}\rangle$  \\ \midrule
        0.6   & 0.01943$^{-0.00161}_{+0.00179}$   & 26.92$^{+0.38}_{-0.68}$   \\
        %0.7 & -0.04282    &    \\
        0.8  & 0.01073$^{-0.0018}_{+0.00162}$   & 27.93$^{+0.09}_{-0.15}$   \\
        %0.9 & -0.04135   &   \\
        1.0 &  0.001938$^{-0.00181}_{+0.00163}$  &  28.28$^{+0.04}_{-0.02}$ \\
        %1.1 & -0.03977  &   \\
        1.2  & -0.00698$^{-0.00184}_{+0.00165}$  & 28.43$^{+0.01}_{-0}$  \\ \midrule
        %1.27 & -0.03835   &   \\
        GX1APN & & \\
        $g^{\rm eff}_{\rm A}$ & sNME$_{\beta^+/\rm EC}$(2) & $\langle E_{\beta^+}\rangle$  \\ \midrule
        %0.818 & 0.02697$^{+0.00164}_{-0.00181}$ & 28.41 & &\\ \midrule
        1.0 & 0.13972$_{+0.00183}^{-0.00165}$& 28.47$_{+0}^{-0}$ \\
        \midrule
        FPD6PN & & \\
        $g^{\rm eff}_{\rm A}$ & sNME$_{\beta^+/\rm EC}$(2) & $\langle E_{\beta^+}\rangle$  \\ \midrule
        1.0 & 0.05715$_{+0.00181}^{-0.00164}$ & 28.64$_{+0.01}^{-0}$ \\
        \bottomrule
    \end{tabular}}
\end{table}

%%\label{}

\section{Summary and conclusions}
To summarize, in this letter a novel and powerful method, namely the Branching Ratio Method (BRM), is introduced for $g_{\rm A}^{\rm eff}$ determination for $\beta^+$/EC decays. As an example, the BRM is applied in modeling the physics of 2$^{nd}$ FNU $\beta^+$/EC decay of $^{59}$Ni. Using the BRM it becomes clear that out of the three Hamiltonians, GX1APN is the best interaction as it is the one that reproduces the experimental value of the branching ratio $\rm BR_{\beta^+: \rm EC}$ of Eq.~(\ref{relbr}) and the experimental total half-life $t_{1/2}^{\beta^+/\rm EC}$ simultaneously.
A powerful aspect of the BRM becomes evident in the case of 2$^{nd}$ FNU $\beta^+$/EC decay of $^{59}$Ni from its ability to differentiate between two sNME solutions that reproduce the (partial) half-lives of Eq.~(\ref{halflife}), since the values of $\rm BR_{\beta^+: \rm EC}$ for the two sNME values are of different orders of magnitude for all the adopted interactions. Such differentiation helps identify the physically relevant sNME making it possible to determine $g_{\rm A}^{\rm eff}$ uniquely, as in the case of the GX1APN interaction. In the present case, the values $g_{\rm A}^{\rm eff}=0.818^{+0.317}_{-0.208}$ and $\textrm{sNME}=0.0270^{+0.0255}_{-0.016}$ are determined with significant uncertainty consequent of the large uncertainties in the (partial) half-lives of Eq.~(\ref{halflife}) to which the two parameters are fitted. 

It would be interesting and important to compare our obtained "default" positron spectral shape with a measured one in order to have a further confirmation of the suitability of the GX1APN interaction for reliable prediction of beta-decay properties of the $pf$-shell nuclei. Furthermore, we could show that the other two interactions, KB3GPN and FPD6PN, studied in this work, have clear shortcomings in this respect. The here studied KB3GPN interaction is a showcase of $g_{\rm A}$-dependent positron spectral shapes and as such amenable to the enhanced SSM (spectrum-shape method) or SMM (spectral-moments method) type of determination of $g_{\rm A}^{\rm eff}$, never conducted in the $\beta^+$/EC side of nuclear $\beta$ decays.
%$g_{\rm A}^{\rm eff}$ determined here for GX1APN interaction using BRM, and $g_{\rm A}^{\rm eff}$ that can in principle be determined using Enhanced SSM and SMM from $g_{\rm A}^{\rm eff}$/sNME dependent spectral shapes of KB3GPN interaction if physically relevant spectral geometries are present amongst spectral shapes corresponding to range of $g_{\rm A}^{\rm eff}$.
The complementary aspects and interplay between the BRM, enhanced SSM, and SMM remain an interesting open question as experiments are necessary for such studies to determine the spectral shape for FNU $\beta^+$/EC decays.
%Findings of Enhanced SSM and SMM concerning physically relevant sNMEs are consistent with the findings of BRM discussed in this letter, in that the existence of unique physically relevant sNME in the presence of two sNME solutions, and consequently unique $g_{\rm A}^{\rm eff}$ for a given Nuclear Hamiltonian.  
The branching-ratio method could usher a renaissance in our understanding of the $\beta^+$/EC decays, while simultaneously testing the suitability and limitations of nuclear Hamiltonians in modeling weak-interaction processes relevant for the neutrinoless double beta decay, nuclear astrophysics, reactor neutrinos and small-decay-energy beta decays.

%%\label{}

\section*{Acknowledgements}
We acknowledge support from project PNRR-I8/C9-CF264, Contract No. 760100/23.5.2023 of the Romanian Ministry of Research, Innovation and Digitization (the NEPTUN project). We acknowledge grants of computer capacity from the Finnish Grid and
Cloud Infrastructure (persistent identifier urn:nbn:fi:research-infras-2016072533 ).

\appendix

%\section{Appendix title 1}

%\section{Appendix title 2}

%\bibliographystyle{elsarticle-num.bst}
%\biboptions{sort&compress}
\bibliography{example}

\begin{thebibliography}{48}
\providecommand{\natexlab}[1]{#1}
\providecommand{\url}[1]{\texttt{#1}}
\expandafter\ifx\csname urlstyle\endcsname\relax
  \providecommand{\doi}[1]{doi: #1}\else
  \providecommand{\doi}{doi: \begingroup \urlstyle{rm}\Url}\fi

\bibitem[NND(2024)]{NNDC}
\textrm{ENSDF} database.
\newblock \emph{\textrm{National Nuclear Data Center (NNDC)}}, 2024.
\newblock URL \url{https://www.nndc.bnl.gov/ensdf/}.

\bibitem[Agnihotri et~al.(2024)Agnihotri, Suhonen, and Kim]{PhysRevLett.133.232501}
Aagrah Agnihotri, Jouni Suhonen, and Hong~Joo Kim.
\newblock Constraints for rare electron-capture decays mimicking detection of dark-matter particles in nuclear transitions.
\newblock \emph{Phys. Rev. Lett.}, 133:\penalty0 232501, Dec 2024.
\newblock \doi{10.1103/PhysRevLett.133.232501}.

\bibitem[Agostini et~al.(2023)Agostini, Benato, Detwiler, Men\'endez, and Vissani]{RevModPhys.95.025002}
Matteo Agostini, Giovanni Benato, Jason~A. Detwiler, Javier Men\'endez, and Francesco Vissani.
\newblock Toward the discovery of matter creation with neutrinoless ${\beta}{\beta}$ decay.
\newblock \emph{Rev. Mod. Phys.}, 95:\penalty0 025002, 2023.

\bibitem[Aker et~al.(2022)Aker, {\textit{et al.}}, and {(The KATRIN Collaboration)}]{NatPhys18}
M.~Aker, {\textit{et al.}}, and {(The KATRIN Collaboration)}.
\newblock Direct neutrino-mass measurement with sub-electronvolt sensitivity.
\newblock \emph{Nat. Phys.}, 18:\penalty0 160–166, 2022.
\newblock \doi{https://doi.org/10.1038/s41567-021-01463-1}.

\bibitem[Bandac et~al.(2024)Bandac, Berg{\' e}, Calvo-Mozota, Carniti, Chapellier, Danevich, Dixon, Dumoulin, Ferri, Giuliani, Gotti, Gras, Helis, Imbert, Khalife, Kobychev, Kostensalo, Loaiza, {de Marcillac}, Marnieros, Marrache-Kikuchi, Martinez, Nones, Olivieri, {Ortiz de Sol{\' o}rzano}, Pessina, Poda, Scarpaci, Suhonen, Tretyak, Zarytskyy, and Zolotarova]{Bandac2024}
I.~Bandac, L.~Berg{\' e}, J.~M. Calvo-Mozota, P.~Carniti, M.~Chapellier, F.~A. Danevich, T.~Dixon, L.~Dumoulin, F.~Ferri, A.~Giuliani, C.~Gotti, Ph. Gras, D.~L. Helis, L.~Imbert, H.~Khalife, V.~V. Kobychev, J.~Kostensalo, P.~Loaiza, P.~{de Marcillac}, S.~Marnieros, C.~A. Marrache-Kikuchi, M.~Martinez, C.~Nones, E.~Olivieri, A.~{Ortiz de Sol{\' o}rzano}, G.~Pessina, D.~V. Poda, J.~A. Scarpaci, J.~Suhonen, V.~I. Tretyak, M.~Zarytskyy, and A.~Zolotarova.
\newblock Precise {$^{113}$Cd} {$\beta$} decay spectral shape measurement and interpretation in terms of possible {$g_A$} quenching.
\newblock \emph{European Physical Journal C}, 84:\penalty0 1158, 2024.

\bibitem[Behrens and B{\"u}hring(1982)]{Behrens1982ElectronRW}
Heinrich Behrens and Wolfgang B{\"u}hring.
\newblock \emph{Electron radial wave functions and nuclear beta-decay}.
\newblock Oxford University Press, 1982.

\bibitem[Blaum et~al.(2020)Blaum, Eliseev, Danevich, Tretyak, Kovalenko, Krivoruchenko, Novikov, and Suhonen]{RevModPhys.92.045007}
K.~Blaum, S.~Eliseev, F.~A. Danevich, V.~I. Tretyak, Sergey Kovalenko, M.~I. Krivoruchenko, Yu.~N. Novikov, and J.~Suhonen.
\newblock Neutrinoless double-electron capture.
\newblock \emph{Rev. Mod. Phys.}, 92:\penalty0 045007, 2020.

\bibitem[Bodenstein-Dresler et~al.(2020)Bodenstein-Dresler, Chu, Gehre, Gößling, Heimbold, Herrmann, Hodak, Kostensalo, Kröninger, Küttler, Nitsch, Quante, Rukhadze, Stekl, Suhonen, Tebrügge, Temminghoff, Volkmer, Zatschler, and Zuber]{2020135092}
Lucas Bodenstein-Dresler, Yingjie Chu, Daniel Gehre, Claus Gößling, Arne Heimbold, Christian Herrmann, Rastislav Hodak, Joel Kostensalo, Kevin Kröninger, Julia Küttler, Christian Nitsch, Thomas Quante, Ekaterina Rukhadze, Ivan Stekl, Jouni Suhonen, Jan Tebrügge, Robert Temminghoff, Juliane Volkmer, Stefan Zatschler, and Kai Zuber.
\newblock Quenching of {$g_{\rm A}$} deduced from the {$\beta$}-spectrum shape of {$^{113}$Cd} measured with the {COBRA} experiment.
\newblock \emph{Physics Letters B}, 800:\penalty0 135092, 2020.
\newblock ISSN 0370-2693.
\newblock \doi{https://doi.org/10.1016/j.physletb.2019.135092}.

\bibitem[Brown and Rae(2014)]{BROWN2014115}
B.A. Brown and W.D.M. Rae.
\newblock \textrm{The Shell-Model Code NuShellX$@$MSU}.
\newblock \emph{Nuclear Data Sheets}, 120:\penalty0 115--118, 2014.

\bibitem[de~Roubin et~al.(2020)de~Roubin, Kostensalo, Eronen, Canete, de~Groote, Jokinen, Kankainen, Nesterenko, Moore, Rinta-Antila, Suhonen, and Vil\'en]{PhysRevLett.124.222503}
A.~de~Roubin, J.~Kostensalo, T.~Eronen, L.~Canete, R.~P. de~Groote, A.~Jokinen, A.~Kankainen, D.~A. Nesterenko, I.~D. Moore, S.~Rinta-Antila, J.~Suhonen, and M.~Vil\'en.
\newblock High-precision {$Q$}-value measurement confirms the potential of $^{135}{Cs}$ for absolute antineutrino mass scale determination.
\newblock \emph{Phys. Rev. Lett.}, 124:\penalty0 222503, 2020.

\bibitem[Ejiri et~al.(2019)Ejiri, Suhonen, and Zuber]{Neutrinonuclearres}
H.~Ejiri, J.~Suhonen, and K.~Zuber.
\newblock Neutrino–nuclear responses for astro-neutrinos, single beta decays and double beta decays.
\newblock \emph{Physics Reports}, 797:\penalty0 1, 2019.

\bibitem[Engel and Menéndez(2017)]{Engel_2017}
Jonathan Engel and Javier Menéndez.
\newblock Status and future of nuclear matrix elements for neutrinoless double-beta decay: a review.
\newblock \emph{Reports on Progress in Physics}, 80\penalty0 (4):\penalty0 046301, 2017.
\newblock \doi{10.1088/1361-6633/aa5bc5}.

\bibitem[Gastaldo et~al.(2017)Gastaldo, Blaum, Chrysalidis, and {\textit{et al.}}]{Gastaldo2017}
L.~Gastaldo, K.~Blaum, K.~Chrysalidis, and {\textit{et al.}}
\newblock The electron capture in {$^{163}$Ho} experiment – {ECHo}.
\newblock \emph{Eur. Phys. J. Spec. Top.}, 226:\penalty0 1623–1694, 2017.

\bibitem[Ge et~al.(2021)Ge, Eronen, Tyrin, Kotila, Kostensalo, Nesterenko, Beliuskina, de~Groote, de~Roubin, Geldhof, Gins, Hukkanen, Jokinen, Kankainen, Koszor\'us, Krivoruchenko, Kujanp\"a\"a, Moore, Raggio, Rinta-Antila, Suhonen, Virtanen, Weaver, and Zadvornaya]{PhysRevLett.127.272301}
Z.~Ge, T.~Eronen, K.~S. Tyrin, J.~Kotila, J.~Kostensalo, D.~A. Nesterenko, O.~Beliuskina, R.~de~Groote, A.~de~Roubin, S.~Geldhof, W.~Gins, M.~Hukkanen, A.~Jokinen, A.~Kankainen, \'A. Koszor\'us, M.~I. Krivoruchenko, S.~Kujanp\"a\"a, I.~D. Moore, A.~Raggio, S.~Rinta-Antila, J.~Suhonen, V.~Virtanen, A.~P. Weaver, and A.~Zadvornaya.
\newblock $^{159}{Dy}$ electron-capture: A new candidate for neutrino mass determination.
\newblock \emph{Phys. Rev. Lett.}, 127:\penalty0 272301, 2021.

\bibitem[Ge et~al.(2022)Ge, Eronen, {de Roubin}, Tyrin, Canete, Geldhof, Jokinen, Kankainen, Kostensalo, Kotila, Krivoruchenko, Moore, Nesterenko, Suhonen, and Vilén]{GE2022137226}
Z.~Ge, T.~Eronen, A.~{de Roubin}, K.S. Tyrin, L.~Canete, S.~Geldhof, A.~Jokinen, A.~Kankainen, J.~Kostensalo, J.~Kotila, M.I. Krivoruchenko, I.D. Moore, D.A. Nesterenko, J.~Suhonen, and M.~Vilén.
\newblock High-precision electron-capture {$Q$} value measurement of {$^{111}$In} for electron-neutrino mass determination.
\newblock \emph{Physics Letters B}, 832:\penalty0 137226, 2022.
\newblock ISSN 0370-2693.
\newblock \doi{https://doi.org/10.1016/j.physletb.2022.137226}.

\bibitem[Gysbers et~al.(2019)Gysbers, Hagen, Holt, Jansen, Morris, Navratil, Papenbrock, Quaglioni, Schwenk, Stroberg, and Wendt]{quenchingresolved}
P.~Gysbers, G.~Hagen, J.~D. Holt, G.~R. Jansen, T.~D. Morris, P.~Navratil, T.~Papenbrock, S.~Quaglioni, A.~Schwenk, S.~R. Stroberg, and K.~A. Wendt.
\newblock Discrepancy between experimental and theoretical $\beta$-decay rates resolved from first principles.
\newblock \emph{Nat. Phys.}, 15:\penalty0 428, 2019.

\bibitem[Haaranen et~al.(2016)Haaranen, Srivastava, and Suhonen]{introssm}
M.~Haaranen, P.~C. Srivastava, and J.~Suhonen.
\newblock Forbidden nonunique ${\beta}$ decays and effective values of weak coupling constants.
\newblock \emph{Phys. Rev. C}, 93:\penalty0 034308, 2016.

\bibitem[Haaranen et~al.(2017)Haaranen, Kotila, and Suhonen]{ssm}
M.~Haaranen, J.~Kotila, and J.~Suhonen.
\newblock Spectrum-shape method and the next-to-leading-order terms of the ${\beta}$-decay shape factor.
\newblock \emph{Phys. Rev. C}, 95:\penalty0 024327, 2017.

\bibitem[Hariasz et~al.(2023)Hariasz, Stukel, Di~Stefano, Rasco, Rykaczewski, Brewer, Stracener, Liu, Gai, Rouleau, Carter, Kostensalo, Suhonen, Davis, Lukosi, Goetz, Grzywacz, Mancuso, Petricca, Fija\l{}kowska, Woli\'{n}ska-Cichocka, Ninkovic, Lechner, Ickert, Morgan, Renne, and Yavin]{PhysRevC.108.014327}
L.~Hariasz, M.~Stukel, P.~C.~F. Di~Stefano, B.~C. Rasco, K.~P. Rykaczewski, N.~T. Brewer, D.~W. Stracener, Y.~Liu, Z.~Gai, C.~Rouleau, J.~Carter, J.~Kostensalo, J.~Suhonen, H.~Davis, E.~D. Lukosi, K.~C. Goetz, R.~K. Grzywacz, M.~Mancuso, F.~Petricca, A.~Fija\l{}kowska, M.~Woli\'{n}ska-Cichocka, J.~Ninkovic, P.~Lechner, R.~B. Ickert, L.~E. Morgan, P.~R. Renne, and I.~Yavin.
\newblock Evidence for ground-state electron capture of $^{40}{K}$.
\newblock \emph{Phys. Rev. C}, 108:\penalty0 014327, Jul 2023.
\newblock \doi{10.1103/PhysRevC.108.014327}.

\bibitem[Hayen et~al.(2019{\natexlab{a}})Hayen, Kostensalo, Severijns, and Suhonen]{Hayen2019a}
L.~Hayen, J.~Kostensalo, N.~Severijns, and J.~Suhonen.
\newblock First-forbidden transitions in reactor antineutrino anomaly.
\newblock \emph{Phys. Rev. C}, 100:\penalty0 054323, 2019{\natexlab{a}}.

\bibitem[Hayen et~al.(2019{\natexlab{b}})Hayen, Kostensalo, Severijns, and Suhonen]{Hayen2019b}
L.~Hayen, J.~Kostensalo, N.~Severijns, and J.~Suhonen.
\newblock First-forbidden transitions in reactor antineutrino spectra.
\newblock \emph{Phys. Rev. C}, 99:\penalty0 031301, 2019{\natexlab{b}}.

\bibitem[Keblbeck et~al.(2023)Keblbeck, Bhandari, Gamage, Horana~Gamage, Leach, Mougeot, and Redshaw]{PhysRevC.107.015504}
D.~K. Keblbeck, R.~Bhandari, N.~D. Gamage, M.~Horana~Gamage, K.~G. Leach, X.~Mougeot, and M.~Redshaw.
\newblock Updated evaluation of potential ultralow {$Q$}-value ${\beta}$-decay candidates.
\newblock \emph{Phys. Rev. C}, 107:\penalty0 015504, 2023.

\bibitem[Kirsebom et~al.(2019)Kirsebom, Hukkanen, Kankainen, Trzaska, Str\"omberg, Mart\'{\i}nez-Pinedo, Andersen, Bodewits, Brown, Canete, Cederk\"all, Enqvist, Eronen, Fynbo, Geldhof, de~Groote, Jenkins, Jokinen, Joshi, Khanam, Kostensalo, Kuusiniemi, Langanke, Moore, Munch, Nesterenko, Ovejas, Penttil\"a, Pohjalainen, Reponen, Rinta-Antila, Riisager, de~Roubin, Schotanus, Srivastava, Suhonen, Swartz, Tengblad, Vilen, V\'{\i}nals, and \"Ayst\"o]{PhysRevC.100.065805}
O.~S. Kirsebom, M.~Hukkanen, A.~Kankainen, W.~H. Trzaska, D.~F. Str\"omberg, G.~Mart\'{\i}nez-Pinedo, K.~Andersen, E.~Bodewits, B.~A. Brown, L.~Canete, J.~Cederk\"all, T.~Enqvist, T.~Eronen, H.~O.~U. Fynbo, S.~Geldhof, R.~de~Groote, D.~G. Jenkins, A.~Jokinen, P.~Joshi, A.~Khanam, J.~Kostensalo, P.~Kuusiniemi, K.~Langanke, I.~Moore, M.~Munch, D.~A. Nesterenko, J.~D. Ovejas, H.~Penttil\"a, I.~Pohjalainen, M.~Reponen, S.~Rinta-Antila, K.~Riisager, A.~de~Roubin, P.~Schotanus, P.~C. Srivastava, J.~Suhonen, J.~A. Swartz, O.~Tengblad, M.~Vilen, S.~V\'{\i}nals, and J.~\"Ayst\"o.
\newblock Measurement of the ${2}^{+}{\rightarrow}{0}^{+}$ ground-state transition in the ${\beta}$ decay of $^{20}{F}$.
\newblock \emph{Phys. Rev. C}, 100:\penalty0 065805, Dec 2019.
\newblock \doi{10.1103/PhysRevC.100.065805}.

\bibitem[Kostensalo et~al.(2021)Kostensalo, Suhonen, Volkmer, Zatschler, and Zuber]{Kostensalo2021}
Joel Kostensalo, Jouni Suhonen, Juliane Volkmer, Stefan Zatschler, and Kai Zuber.
\newblock Confirmation of {$g_{\rm A}$} quenching using the revised spectrum-shape method for the analysis of the {$^{113}$Cd} {$\beta$}-decay as measured with the {COBRA} demonstrator.
\newblock \emph{Physics Letters B}, 822:\penalty0 136652, 2021.
\newblock ISSN 0370-2693.
\newblock \doi{https://doi.org/10.1016/j.physletb.2021.136652}.

\bibitem[Kostensalo et~al.(2023)Kostensalo, Lisi, Marrone, and Suhonen]{Kostensalo2023}
Joel Kostensalo, Eligio Lisi, Antonio Marrone, and Jouni Suhonen.
\newblock $^{113}{Cd}$ ${\beta}$-decay spectrum and {${g}_{\rm A}$} quenching using spectral moments.
\newblock \emph{Phys. Rev. C}, 107:\penalty0 055502, May 2023.
\newblock \doi{10.1103/PhysRevC.107.055502}.

\bibitem[Kostensalo et~al.(2024)Kostensalo, Lisi, Marrone, and Suhonen]{Kostensalo2024}
Joel Kostensalo, Eligio Lisi, Antonio Marrone, and Jouni Suhonen.
\newblock Analysis of $^{115}$in $\beta$ decay through the spectral moment method.
\newblock \emph{Physical Review C}, 110:\penalty0 045503, 2024.

\bibitem[Kumar et~al.(2020)Kumar, Srivastava, Kostensalo, and Suhonen]{24na36clssm}
Anil Kumar, Praveen~C. Srivastava, Joel Kostensalo, and Jouni Suhonen.
\newblock Second-forbidden nonunique ${{\beta}}^{{-}}$ decays of $^{24}{Na}$ and $^{36}{Cl}$ assessed by the nuclear shell model.
\newblock \emph{Phys. Rev. C}, 101:\penalty0 064304, Jun 2020.
\newblock \doi{10.1103/PhysRevC.101.064304}.

\bibitem[Kumar et~al.(2021)Kumar, Srivastava, and Suhonen]{59fe60fessm}
Anil Kumar, Praveen~C Srivastava, and Jouni Suhonen.
\newblock Second-forbidden nonunique $\beta^-$ decays of {$^{59,60}$Fe}: possible candidates for {$g_{\rm A}$} sensitive electron spectral-shape measurements.
\newblock \emph{The European Physical Journal A}, 57:\penalty0 225, 2021.
\newblock ISSN 1434-601X.
\newblock \doi{10.1140/epja/s10050-021-00540-6}.

\bibitem[Langanke and Mart\'{\i}nez-Pinedo(2003)]{RevModPhys.75.819}
K.~Langanke and G.~Mart\'{\i}nez-Pinedo.
\newblock Nuclear weak-interaction processes in stars.
\newblock \emph{Rev. Mod. Phys.}, 75:\penalty0 819--862, Jun 2003.
\newblock \doi{10.1103/RevModPhys.75.819}.

\bibitem[Langanke et~al.(2021)Langanke, Martínez-Pinedo, and Zegers]{Langanke_2021}
K~Langanke, G~Martínez-Pinedo, and R~G~T Zegers.
\newblock Electron capture in stars.
\newblock \emph{Reports on Progress in Physics}, 84\penalty0 (6):\penalty0 066301, may 2021.

\bibitem[Leder et~al.(2022)Leder, Mayer, Ouellet, Danevich, Dumoulin, Giuliani, Kostensalo, Kotila, de~Marcillac, Nones, Novati, Olivieri, Poda, Suhonen, Tretyak, Winslow, and Zolotarova]{PhysRevLett.129.232502}
A.~F. Leder, D.~Mayer, J.~L. Ouellet, F.~A. Danevich, L.~Dumoulin, A.~Giuliani, J.~Kostensalo, J.~Kotila, P.~de~Marcillac, C.~Nones, V.~Novati, E.~Olivieri, D.~Poda, J.~Suhonen, V.~I. Tretyak, L.~Winslow, and A.~Zolotarova.
\newblock Determining ${g}_{A}/{g}_{V}$ with high-resolution spectral measurements using a {${{LiInSe}}_{2}$} bolometer.
\newblock \emph{Phys. Rev. Lett.}, 129:\penalty0 232502, Dec 2022.
\newblock \doi{10.1103/PhysRevLett.129.232502}.

\bibitem[Pagnanini et~al.(2024)Pagnanini, Benato, Carniti, Celi, Chiesa, Corbett, Dafinei, Di~Domizio, Di~Stefano, Ghislandi, Gotti, Helis, Knobel, Kostensalo, Kotila, Nagorny, Pessina, Pirro, Pozzi, Puiu, Quitadamo, Sisti, Suhonen, and Kuznetsov]{PhysRevLett.133.122501}
L.~Pagnanini, G.~Benato, P.~Carniti, E.~Celi, D.~Chiesa, J.~Corbett, I.~Dafinei, S.~Di~Domizio, P.~Di~Stefano, S.~Ghislandi, C.~Gotti, D.~L. Helis, R.~Knobel, J.~Kostensalo, J.~Kotila, S.~Nagorny, G.~Pessina, S.~Pirro, S.~Pozzi, A.~Puiu, S.~Quitadamo, M.~Sisti, J.~Suhonen, and S.~Kuznetsov.
\newblock Simultaneous measurement of the half-life and spectral shape of $^{115}{In}$ ${\beta}$ decay with an indium iodide cryogenic calorimeter.
\newblock \emph{Phys. Rev. Lett.}, 133:\penalty0 122501, Sep 2024.
\newblock \doi{10.1103/PhysRevLett.133.122501}.

\bibitem[Paulsen et~al.(2024)Paulsen, Ranitzsch, Loidl, Rodrigues, Kossert, Mougeot, Singh, Leblond, Beyer, Bockhorn, Enss, Wegner, Kempf, and N\"ahle]{Paulsen2024}
M.~Paulsen, P.~{C.-O.} Ranitzsch, M.~Loidl, M.~Rodrigues, K.~Kossert, X.~Mougeot, A.~Singh, S.~Leblond, J.~Beyer, L.~Bockhorn, C.~Enss, M.~Wegner, S.~Kempf, and O.~N\"ahle.
\newblock High precision measurement of the {$^{99}$Tc} {$\beta$} spectrum.
\newblock \emph{Physical Review C}, 110:\penalty0 055503, 2024.

\bibitem[Ramalho and Suhonen(2024{\natexlab{a}})]{PhysRevC.109.014326}
M.~Ramalho and J.~Suhonen.
\newblock Computed total ${\beta}$-electron spectra for decays of {Pb} and {Bi} in the $^{220,222}{Rn}$ radioactive chains.
\newblock \emph{Phys. Rev. C}, 109:\penalty0 014326, Jan 2024{\natexlab{a}}.
\newblock \doi{10.1103/PhysRevC.109.014326}.

\bibitem[Ramalho et~al.(2022{\natexlab{a}})Ramalho, Ge, Eronen, Nesterenko, Jaatinen, Jokinen, Kankainen, Kostensalo, Kotila, Krivoruchenko, Suhonen, Tyrin, and Virtanen]{PhysRevC.106.015501}
M.~Ramalho, Z.~Ge, T.~Eronen, D.~A. Nesterenko, J.~Jaatinen, A.~Jokinen, A.~Kankainen, J.~Kostensalo, J.~Kotila, M.~I. Krivoruchenko, J.~Suhonen, K.~S. Tyrin, and V.~Virtanen.
\newblock Observation of an ultralow-{$Q$}-value electron-capture channel decaying to $^{75}{As}$ via a high-precision mass measurement.
\newblock \emph{Phys. Rev. C}, 106:\penalty0 015501, 2022{\natexlab{a}}.

\bibitem[Ramalho et~al.(2022{\natexlab{b}})Ramalho, Suhonen, Kostensalo, Alcal\'a, Algora, Fallot, Porta, and Zakari-Issoufou]{Ramalho2022}
M.~Ramalho, J.~Suhonen, J.~Kostensalo, G.~A. Alcal\'a, A.~Algora, M.~Fallot, A.~Porta, and A.-A. Zakari-Issoufou.
\newblock Analysis of the total ${\beta}$-electron spectrum of {$^{92}{Rb}$}: Implications for the reactor flux anomalies.
\newblock \emph{Phys. Rev. C}, 106:\penalty0 024315, 2022{\natexlab{b}}.

\bibitem[Ramalho et~al.(2024)Ramalho, Suhonen, Neacsu, and Stoica]{Ramalho2024}
M.~Ramalho, J.~Suhonen, A.~Neacsu, and S.~Stoica.
\newblock Spectral shapes of second-forbidden single-transition nonunique {$\beta$} decays assessed using the nuclear shell model.
\newblock \emph{Frontiers of Physics}, 12:\penalty0 1455778, 2024.

\bibitem[Ramalho and Suhonen(2024{\natexlab{b}})]{PhysRevC.109.034321}
Marlom Ramalho and Jouni Suhonen.
\newblock {$g_{\rm A}$}-sensitive {$\beta$} spectral shapes in the mass {$A=86-99$} region assessed by the nuclear shell model.
\newblock \emph{Phys. Rev. C}, 109:\penalty0 034321, Mar 2024{\natexlab{b}}.
\newblock \doi{10.1103/PhysRevC.109.034321}.

\bibitem[Redshaw(2023)]{Redshaw2023}
M.~Redshaw.
\newblock Precise {$Q$} value determinations for forbidden and low energy {$\beta$}-decays using penning trap mass spectrometry.
\newblock \emph{Eur. Phys. J. A}, 59:\penalty0 18, 2023.

\bibitem[Ruotsalainen et~al.(2024)Ruotsalainen, Stryjczyk, Ramalho, Eronen, Ge, Kankainen, Mougeot, and Suhonen]{Ruotsalainen:2024gxu}
J.~Ruotsalainen, M.~Stryjczyk, M.~Ramalho, T.~Eronen, Z.~Ge, A.~Kankainen, M.~Mougeot, and J.~Suhonen.
\newblock Ultra-low $q_\beta$ value for the allowed decay of $^{110}$ag$^m$ confirmed via mass measurements, 9 2024.

\bibitem[Schoppmann(2021)]{universe7100360}
Stefan Schoppmann.
\newblock Status of anomalies and sterile neutrino searches at nuclear reactors.
\newblock \emph{Universe}, 7\penalty0 (10):\penalty0 360, 2021.
\newblock ISSN 2218-1997.
\newblock \doi{10.3390/universe7100360}.
\newblock URL \url{https://www.mdpi.com/2218-1997/7/10/360}.

\bibitem[Suhonen(2007)]{Suhonen2007}
Jouni Suhonen.
\newblock \emph{From Nucleons to Nucleus: Concepts of Microscopic Nuclear Theory}.
\newblock Springer Berlin, Heidelberg, 2007.

\bibitem[Suhonen(2017{\natexlab{a}})]{PhysRevC.96.055501}
Jouni Suhonen.
\newblock Impact of the quenching of {${g}_{\rm A}$} on the sensitivity of $0{\nu}{\beta}{\beta}$ experiments.
\newblock \emph{Phys. Rev. C}, 96:\penalty0 055501, 2017{\natexlab{a}}.

\bibitem[Suhonen and Kostensalo(2019)]{Suhonen2019}
Jouni Suhonen and Joel Kostensalo.
\newblock Double $\beta$ decay and the axial strength.
\newblock \emph{Frontiers in Physics}, 7:\penalty0 29, 2019.
\newblock ISSN 2296-424X.

\bibitem[Suhonen(2017{\natexlab{b}})]{avcreview}
Jouni~T. Suhonen.
\newblock \textrm{Value of the Axial-Vector Coupling Strength in $\beta$ and $\beta\beta$ Decays: A Review}.
\newblock \emph{Frontiers in Physics}, 5:\penalty0 55, 2017{\natexlab{b}}.

\bibitem[Surender et~al.(2024)Surender, Kumar, and Srivastava]{SURENDER2024169772}
Surender, Vikas Kumar, and Praveen~C. Srivastava.
\newblock Study of {$\beta^+$/EC}-decay properties of sd shell nuclei using nuclear shell model.
\newblock \emph{Annals of Physics}, 470:\penalty0 169772, 2024.
\newblock ISSN 0003-4916.
\newblock \doi{https://doi.org/10.1016/j.aop.2024.169772}.

\bibitem[Ydrefors et~al.(2010)Ydrefors, Mustonen, and Suhonen]{ECformalism}
E.~Ydrefors, M.T. Mustonen, and J.~Suhonen.
\newblock \uppercase{MQPM} description of the structure and beta decays of the odd \uppercase{A}=95,97 \uppercase{M}o and \uppercase{T}c isotopes.
\newblock \emph{Nuclear Physics A}, 842\penalty0 (1):\penalty0 33--47, 2010.
\newblock ISSN 0375-9474.
\newblock \doi{https://doi.org/10.1016/j.nuclphysa.2010.04.005}.

\bibitem[Zhang et~al.(2024)Zhang, Qian, and Fallot]{ZHANG2024104106}
Chao Zhang, Xin Qian, and Muriel Fallot.
\newblock Reactor antineutrino flux and anomaly.
\newblock \emph{Progress in Particle and Nuclear Physics}, 136:\penalty0 104106, 2024.

\end{thebibliography}

%\bibliography{example}



%% For example:

%% \bibitem[Aladro et al.(2015)]{Aladro15} Aladro, R., Martín, S., Riquelme, D., et al. 2015, \aas, 579, A101

%%\end{thebibliography}

\end{document}